# Filtering Patent Maps for Visualization of Diversification Paths of Inventors and Organizations

**Bowen Yan**

SUTD-MIT International Design Centre
& Engineering Product Development Pillar
Singapore University of Technology and Design
8 Somapah Road, Singapore 487372
Email: bowen_yan@sutd.edu.sg

**Jianxi Luo**[*]

SUTD-MIT International Design Centre
& Engineering Product Development Pillar
Singapore University of Technology and Design
8 Somapah Road, Singapore 487372
Email: luo@sutd.edu.sg

[*] Corresponding author

# Abstract

In the information science literature, recent studies have used patent databases and patent classification information to construct network maps of patent technology classes. In such a patent technology map, almost all pairs of technology classes are connected, whereas most of the connections between them are extremely weak. This observation suggests the possibility of filtering the patent network map by removing weak links. However, removing links may reduce the explanatory power of the network on inventor or organization diversification. The network links may explain the patent portfolio diversification paths of inventors and inventing organizations. We measure the diversification explanatory power of the patent network map, and present a method to objectively choose an optimal trade-off between explanatory power and removing weak links. We show that this method can remove a degree of arbitrariness compared with previous filtering methods based on arbitrary thresholds, and also identify previous filtering methods that created filters outside the optimal trade-off. The filtered map aims to aid in network visualization analyses of the technological diversification of inventors, organizations and other innovation agents, and potential foresight analysis. Such applications to a prolific inventor (Leonard Forbes) and company (Google) are demonstrated.

# 1 Introduction

Recent studies have proposed to represent the space of technologies as a network (Kay et al., 2014; Leydesdorff et al., 2014; Nakamura et al., 2014; Alstott et al., 2016). Using patent data, the nodes of the network are categories of patents to approximately represent technology fields (e.g. "organic chemistry"), and the links between the nodes are weighted according to the knowledge distance (or proximity) of the technology fields (Engelsman and van Raan, 1994; Joo and Kim, 2010; Yan and Luo, 2016). Such technology networks have been used to analyze the patent portfolio diversification of firms and regions across different technology fields, which are represented as movements in the technology space (Breschi et al., 2003; Jaffe and Trajtenberg, 2002; Rigby, 2013; Kogler et al., 2013; Boschma et al., 2015). Such studies have generally shown some evidence that firms or regions tend to diversify into new technology fields that are proximate to their existing fields (Teece et al., 1994; Breschi et al., 2003; Nesta and Dibiaggio, 2005).

In this paper, we aim to develop a proper technology network map whose links can well explain or predict the likelihoods of diversification of innovation agents across the links. Such a network can be used for visualizing and analyzing the diversification paths of individual innovation agents. The patent technology network is almost fully connected, but the majority of connections are extremely weak compared to a minority of strong connections. In comparison to the stronger links, many weak links could be filtered out to reveal a clearer network structure, as suggested in earlier studies of science maps or patent maps (Klavans and Boyack, 2006a; Leydesdorff et al., 2014). The filtered network structure could be more effective for the visualization and at the same time maintain its effectiveness for the analysis of technological diversification of regions or organizations (Hidalgo et al., 2007). However, which links, and how many, should be removed is unclear.

Therefore, we present a method to filter the technology network based on two principles. The first principle is to remove as many weak links as possible. The second principle is to maintain the overall power of the filtered network to explain the diversification paths of innovation agents (e.g., inventors, R&D organizations) across technology fields throughout the network. However, these two principles conflict, because removing network links will most likely reduce the network's explanatory power. A trade-off must be made, to reconcile this conflict and derive heavily filtered technology networks that maintain high explanatory power on diversification patterns of innovation agents.

Our method is specifically developed to filter the original technology networks for the interests of capturing cross-field diversification paths of two types of innovation agents, i.e. inventors and R&D organizations. We will show that, for different types of innovation agents, filtering thresholds are chosen differently. Alternative network filtering techniques from the literature and this paper are compared. The results show that our method can remove a degree of arbitrariness compared with previous filtering methods based on arbitrary thresholds, and also identify previous filtering methods that created filters outside the optimal trade-off. The filtered maps aim to aid in more effective network visualization analyses of the technological diversification paths of inventors, R&D firms and organizations, and other innovation agents. The applications to a prolific inventor (Leonard Forbes) and company (Google) are demonstrated.

## 2. Literature Review

*2.1 Technological Distance and Diversification*

Prior empirical studies have shown that firms and regions are more likely to diversify across technology areas with high knowledge proximity. For instance, by examining the patterns of technological diversification of the firms in a few most developed countries, according to their records of patenting to the European Patent Office from 1982 to 1993, Breschi et al. (2003) found knowledge proximity of technology fields is a key factor to determine firms' technological diversification across fields. They used co-classification codes contained in patent documents to measure knowledge proximity between technology classes. At the city level, Rigby (2013) showed that US cities' entries into and exits from technology fields are highly related to the level of knowledge proximity among technology fields. He used USPC (United States Patent Classification) classes to represent technology fields, and measured inter-field proximity as the probability that a patent in class *j* will cite a patent in class *i*. Boschma et al. (2015) also found similar evidence that the probability for U.S. cities to enter a new technology field is significantly related to its level of knowledge proximity with existing technology fields that the city has entered.

In general, these studies have suggested that the network maps of patent technology classes and measures of distance or proximity among these classes can be used to explain or predict the likelihoods of diversification of firms and regions across pairs of technology classes.

*2.2 Technology Network Maps and Filtering*

Several recent studies have constructed and visualized filtered patent technology networks. For instance, Leydesdorff et al. (2014) constructed a technology network map, using IPC classes of USPTO patents as nodes. The weight of a link between two technology classes is measured using cosine similarity (Jaffe, 1986), i.e. the angular cosine value of the two vectors of citations from the patents of these two classes to all other classes respectively. When visualizing the network and detecting its community structure in Pajek, Leydesdorff et al. (2014) suggested that, "*without a threshold for the cosine, the visualization is not informative.*" However, they only reported the community detection results, given the threshold value of cosine > 0.2. That is, inter-field links with weight value <= 0.2 were removed. They did not discuss why this specific threshold value was chosen.

Kay et al. (2014) also used the cosine similarity to measure the "technology distance", i.e. link weight, among different IPC technology classes in their network. They aggregated or decomposed the original IPC patent classes into equally sized categories, to optimize the size distribution of the nodes in the network. They also found that the technology network is highly interconnected. Their network visualizations using Pajek included all the links regardless of cosine values, and appear to be highly dense and not informative.

Klavans and Boyack (2006a) proposed a method to create and visualize very large maps of hundreds of thousands of scientific papers, using a modified cosine measure. They filtered their large dense networks, by adding links in the order of decreasing weights back to the network, until all unique nodes are connected into the network for the first time. Their technique is equivalent to removing the weakest links in the order of increasing weights, until the removal of one additional stronger link would cause the network to become not fully connected.

Hidalgo et al. (2007) constructed and visualized the network of product categories based on international trade data. Different product categories are connected according to a measure of proximity. The proximity is calculated as the likelihood for an average country to develop strong relative comparative advantage (RCA) in one product category, given that it has developed strong RCA in the other. The assumption is that this likelihood is high if the capabilities required to produce products in one category are similar to those required to produce another product. Their network of product categories is almost fully connected and thus super dense. To offer a simple network visualization, they superposed links with a value>0.55 on the maximum spanning tree (MST). A MST has the minimal set of strongest links that keep the network connected (i.e., MST has specifically $n$-1 links, $n$ is the total number of nodes in the network). Their choice of the threshold of 0.55 is determined based on the "rule of thumb" that good network visualization has an average degree equal to 4 (or the number of links is twice of the number of nodes). However, the impact of this network filtering for visualization ease on the utility of the network to predict cross-field diversification is not clear.

In sum, the literature has suggested that weighted patent technology networks can be used to explain the technology diversification paths of innovation agents, such as firms and cities. In the meantime, such networks are densely connected, and need to be filtered to improve visualization and leave a clear network structure for analysis and sense making. In particular, the patent network mapping efforts reviewed above all conducted network filtering, and their filtering strategies shared the preference to keep the strongest links and remove a fraction of the weakest links from the original network. (The patterns shown in our Figure 1 and Figure 6 in section 4 provide additional rationales behind such a filtering strategy for our interest in the present paper.)

While these studies used different techniques and strategies for different reasons and needs, in order to filter a network as a holistic whole with valuing stronger links more than weaker ones, a threshold of link weight has always been required to determine the links with values weaker than the threshold for removal. That is, the threshold provides a stopping criterion of link removal. However, in the literature, the threshold was often chosen either arbitrarily or for the sake of graphical visualization. In this study, we will present a method to decide the threshold of network filtering for the interest of using the technology network for visual analysis of the diversification paths of different innovation agents.

## 3. The Technology Network

### 3.1 Diversification Explanatory Power

Innovation studies have suggested that new technological knowledge and capabilities of innovation agents, e.g. inventors, firms or nations, are generally learned or built up incrementally via an evolutionary path, in which their future knowledge positions are shaped by the past ones (Winter, 2000; Nelson and Winter, 2009; Luo et al., 2014). Learning theories (Winston, 1992) also suggested that it will be easier for individuals and organizations to learn and enter new fields that are proximate to the ones that have already understood, because the required knowledge and capabilities for innovation in the current and new fields are highly related or similar to each other (Breschi et al., 2003; Wuyts et al., 2005; Nootebooma et al., 2007). Therefore, one can expect the diversification paths of innovation agents to be highly correlated with the network paths in the technology proximity network (Breschi et al., 2003; Nesta and Dibiaggio, 2005).

Based on this understanding, a patent network's *diversification explanatory power,* i.e., how well the network links serve as the paths of diversification of innovation agents across technology fields (i.e., nodes) in the network, can be assessed as the Pearson correlation coefficient between the knowledge proximity, i.e. weights of the links, of all pairs of technology fields, and the likelihoods that innovation agents diversify across the respective pairs of fields. On a network map whose links well explain diversification paths of an innovation agent, the innovation agent is more likely to diversify across proximate fields. Note that, the explanatory power corresponds to the types (or aggregation levels) of innovation agents of interest. In this paper, we focus on individual inventors and inventive organizations (which can be technology firms, universities or R&D organizations), whereas the same analysis can be conducted for cities, regions and countries, etc.

The explanatory power of a filtered map on inventor and organization diversification patterns needs to be assessed respectively, because inventors and organizations present different likelihoods to diversify across the same pairs of technology fields. The empirical calculations of the cross-field diversification likelihoods of inventors and organizations use different information in the patent data. Specifically, we calculate "inventor diversification likelihood" as the minimum of the pairwise conditional probabilities of an inventor having a patent in one technology class, given that this person also has patents in the other class in a given time period. To calculate this measure, we used the unique inventor identifiers from the Institute for Quantitative Social Science at Harvard University (Li et al., 2014). "Organization diversification likelihood" is calculated as the minimum of the pairwise conditional probabilities of an organization having a patent in one technology class, given that it also has patents in the other in a given time period. To calculate this measure, we used the unique assignee identifiers created by the National Bureau of Economics Research (NBER) (Hall et al., 2001).

*3.2 Constructing the Patent Technology Network*

We used the patent database of USPTO (United States Patent and Trademark Office) to construct the patent technology network. IPC (International Patent Classification) classes are used to represent technological fields as network nodes, following prior patent map studies that have considered IPC classes the most suitable and stable representations of technology fields (Leydesdorff et al., 2014).[1] There are 121 meaningful IPC3 classes, which contain 3,911,054 US patents from 1976 to 2010, after removing a few undefined classes (e.g., "A99 – subject matter not otherwise provided for in this section"). The network of IPC3 classes (3-digit IPC classes) provides the best resolution for technology network visualization without losing necessary details (Yan and Luo, 2016), whereas the network of IPC4 classes (4-digit IPC classes) contains 629 classes as nodes and appears to be visually complex and not informative.

---

[1] We are aware of a few recent publications that proposed hybrid categories of patents, beyond either IPC or USPC, to represent technologies or technology fields, following different principles and objectives. For instance, Kay et al. (2014) assigned patents more evenly across patent categories in order to have better visualization. Benson and Magee (2013;

To quantify link weights in the network, a measure of the "proximity" between pairs of technology classes is required. Proximity is a reverse concept of "distance". Alternative proximity measures exist in the literature (Joo and Kim, 2010; Klavans and Boyack, 2006b). Cosine similarity was a popularly used measure of proximity among different patent classes (Leydesdorff et al., 2014; Kay et al. 2014). A recent comparative analysis of 12 proximity measures for patent maps suggested that Jaccard index is a superior representative choice of alternative ones, because the patent network map resulting from Jaccard index (Jaccard, 1901) is the most correlated with other maps from all other proximity measures (Yan and Luo, 2016). For our interest in the analysis of agent diversification across fields, Jaccard index as a relational measure will be more meaningful than a similarity measure, e.g., cosine similarity.

Jaccard Index is calculated as the number of shared references of the patents in a pair of classes normalized by the number of unique references of patents in either class. Its mathematical form is $\frac{|C_i \cap C_j|}{|C_i \cup C_j|}$, where $C_i$ and $C_j$ are the numbers of references of the patents in technology classes $i$ and $j$; $|C_i \cap C_j|$ is the number of patents referenced in both technology classes $i$ and $j$, and $|C_i \cup C_j|$ is the total number of unique patents referenced in either technology classes $i$ or $j$, respectively. Thus, the index value is between [0,1], and indicates the proximity of the knowledge bases of a pair of technology fields defined by IPC classes. Higher proximity of a pair of fields may give rise to the ease and thus higher likelihood for innovation agents to diversify across them, because much of the knowledge required for innovation in one field can be also used for innovation in the other field.

---

2015) proposed alternative hybrid classes of technologies that aim to better represent actual technology fields with improved completeness and relevancy from a technologist perspective. In the present study, our focus is in a technology network filtering method. Such a method can be applied to networks of alternatively defined patent classes or categories.

We conducted a comparative analysis on the *diversification explanatory powers* of alterative networks constructed using IPC3 or IPC4 classes to represent nodes, and using Jaccard index or cosine similarity to quantify link weights. Table 1 reports the diversification explanatory powers, i.e. the correlation between link weights and cross-link diversification likelihoods for an average inventor or organization, for different combinations. Regardless of IPC3 or IPC4, the networks using Jaccard index present *much higher* explanatory powers on both inventor and organization diversification, than those using cosine similarity. Regardless of Jaccard index or cosine similarity, the networks using IPC3 have *slightly higher* explanatory powers on both inventor and organization diversification, than those using IPC4. Taken together, the network constructed using IPC3 and Jaccard index achieves the highest explanatory power on both inventor and organization diversification.[2] In our later analysis, we will focus on this network for filtering and case analyses.

**Table 1.** Diversification Explanatory Powers of Alternative Patent Technology Networks

|  | Inventor Diversification | | Organization Diversification | |
| --- | --- | --- | --- | --- |
|  | IPC3 | IPC4 | IPC3 | IPC4 |
| Jaccard Index | ***0.93*** | 0.92 | ***0.65*** | 0.57 |
| Cosine Similarity | 0.41 | 0.45 | 0.23 | 0.30 |

Figure 1 shows the correlations between cross-link diversification likelihoods for an average inventor or organization, and link weights in the network constructed using IPC3 and Jaccard index. Several patterns are noteworthy. First, the proximity-diversification correlation is generally stronger for individual inventors than organizations. This difference implies that cross-field diversification and related decisions are less constrained by inter-field knowledge distance for organizations than for individuals. A possible explanation is that organizations can invest and acquire knowledge more flexibly than individual inventors, whereas inventors do not have a wide scope of resources and must learn and master relevant knowledge of a technology field in order to invent there. Such a difference in resources and capacities of organizations and individuals may have also resulted in our second observation in Figure 1 – when knowledge proximity is lower, implying stronger constraints to diversification, inventors are clearly less prone to diversify than organizations. The differences in cross-field diversification likelihoods of inventors and organizations suggest the filtering of them needs to be done differently.

---

[2] For a robustness check, we also compared the diversification explanatory powers of the four networks filtered to different extents, i.e., keeping the same portion of the strongest links in the original networks after weak link removals. Across different extents of filtering, the network using IPC3 and Jaccard index constantly provides the best diversification explanatory power.

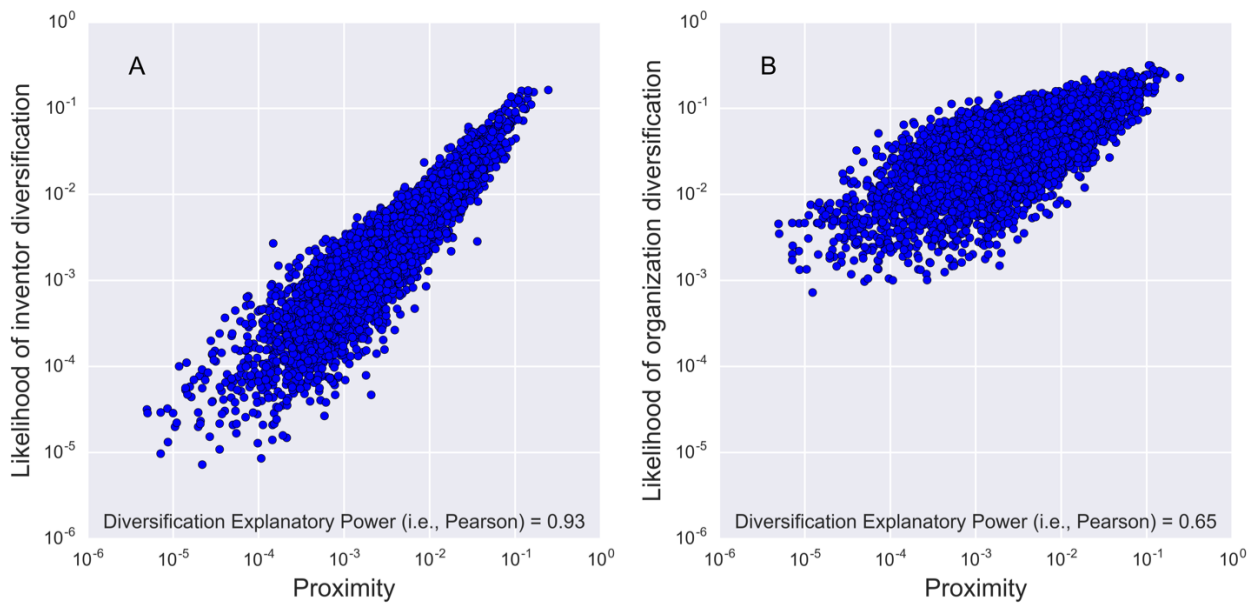

**Figure 1. The technology network using IPC3 and Jaccard index mirrors the likelihoods of inventor and organization diversification across fields.** Each dot represents a pair of technology classes. Link weights and diversification likelihoods are calculated based on all patent records from 1976 to 2010. [3]

In addition, for both organizations and inventors, the scattered dots are bounded in a "dagger" area pointing to the upper right. This pattern further suggests that extremely high knowledge proximity can enable organizations and individuals to diversify similarly, resulting in smaller variance in diversification behaviors. On the other hand, the same pattern also suggests that link weights are more correlated with diversification likelihoods (i.e., leading to higher explanatory power), when link weights are higher. Therefore, when removing links, one should target at weak links for removal first for the sake of keeping its high explanatory power as high as possible. The strategy of removing weaker links first was commonly adopted in the filtering exercises in the literature (Klavans and Boyack 2006a; Hidalgo et al. 2007; Leydesdorff et al. 2014).

---

[3] For a robustness check, we also compared the diversification explanatory powers of alternative networks constructed using data in shorter time periods. The network using IPC3 and Jaccard index has consistently the highest diversification explanatory power.

*3.3 Structure of the Technology Network*

Figure 2 visualizes the full technology network constructed using IPC3 and Jaccard index. It has 7,195 none-zero links. Among the total 121×120/2=7,260 pairs of technology classes, only 65 or 0.9% of them have zero link weight and no direct connection. Thus, the network is almost fully connected and extremely dense.

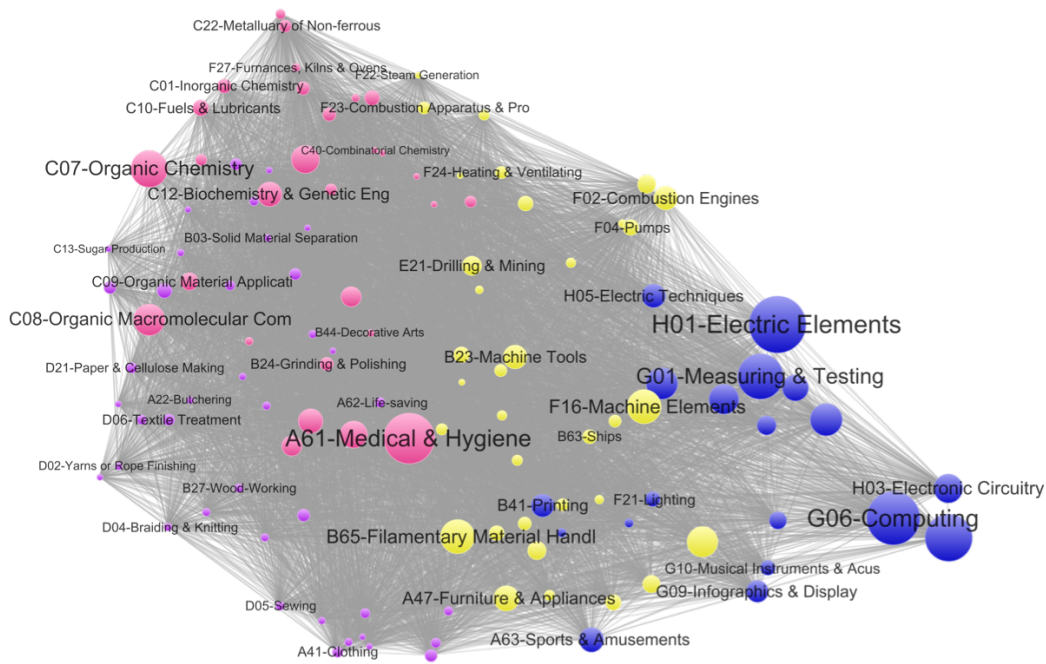

**Figure 2.** The full technology network. Several clusters of technology classes were identified using the Louvain community detection method (Blondel et al., 2008) and colored accordingly.

Figure 3 is the matrix visualization of the same network. Each row and column of the matrix denotes a technology class, and the brightness level of the entry in row *i* and column *j* denotes the knowledge proximity value between technology classes *i* and *j*. The rows and columns, representing technology classes, are sorted in the same order using the "average linkage clustering" algorithm to reveal several relatively dense areas in the network. Most importantly, the matrix visualization clearly reveals very large sparse areas (in dark blue color), in which technology classes are only weakly connected with one another, despite the network's almost fully connected status shown in Figure 2. That is, most pairs of technology classes are only weakly related.

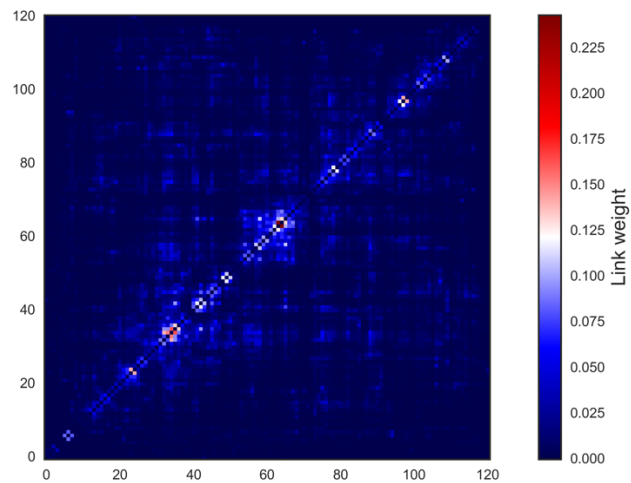

**Figure 3.** Matrix representation of the technology network.

Figure 4 shows the distribution of links by their weights. The distribution is highly skewed toward low link values, i.e., the left). This shape of the distribution curve confirms that in the network, the majority of links are extremely weak, with only a small number of strong links. Out of the total 7,260 theoretically possible links among 121 technology classes, 0.9% of them are equal to zero, 82.5% of them have weights lower than 0.01, 99.7% of them have weights below 0.1, and the maximum weight is 0.243.

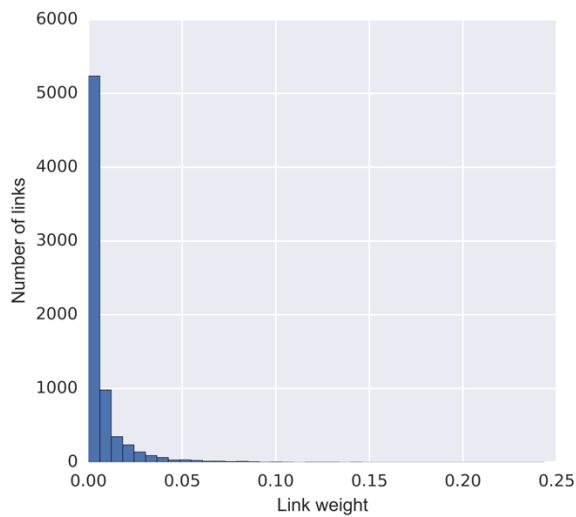

**Figure 4.** Distribution of links by weight.

Therefore, the technology class network is almost fully connected; but most links are extremely weak. The weights of the majority of links are a few orders of magnitude weaker than a small number of relatively strong links. This fact suggests the possibility to simplify the network by removing many of the weakest links. But, which and how many links should be removed? What is the threshold to cut off the links with values below it? In the following, we introduce our method to determine the threshold value, and the stopping criterion for link removal.

**4. Network Filtering**

The purpose of network filtering is to remove as many weak links as possible. But link removal must be done under the constraint that the explanatory power of the filtered network needs to be maintained as high as possible. The main utility of the technology network for our purposes is to explain or predict the likelihoods of inventor and organization diversification across the links between pairs of technology fields according to respective link weights. Removing too many links may reduce the explanatory power of the filtered network. However, it is possible that the network's explanatory power can be improved when removing the weakest links that introduce noise. We will show such a case later. It is also possible that the explanatory power can remain stable or decrease only slightly when certain amount of negligible links is removed.

In the following, we present the procedure to identify the link removal threshold for filtering networks so that they effectively explain the diversification patterns of inventors and organizations across network links.[4] Our filtering method starts with identifying the maximum spanning tree (MST) of the original full technology network, and then incrementally adds the rest of links, one by one, in the order of decreasing weights to the MST. Every time when a link is added back, we calculate the explanatory power of the updated network. When the desired explanatory power is reached for the first time, the filtered network at this point contains the smallest number of links needed to obtain that explanatory power.

---

[4] Herein, we consider and analyze the network as a holistic whole, so a single threshold value is applied across the entire network. However, if one's focus is in individual and separate local communities (i.e. sub-graphs) of the network, different filtering criteria and threshold values can be applied in different separate local communities within the network.

In particular, starting with MST as the backbone guarantees that all nodes are connected in the network, forming a full technology space map. A full map of all technological fields will be useful for comparative and longitudinal analyses of the knowledge positions and diversification paths innovation agents. For instance, different inventors and organizations with patents in different fields can be shown on the same full map simultaneously, so their differentiated locations can be revealed. For a single innovation agent that diversifies into new fields over time, his/her/its old and new fields and the network links connecting them can be simultaneously visualized on the same map, revealing the diversification paths of the innovation agent over time. Two examples of such applications are demonstrated in section 5.1.

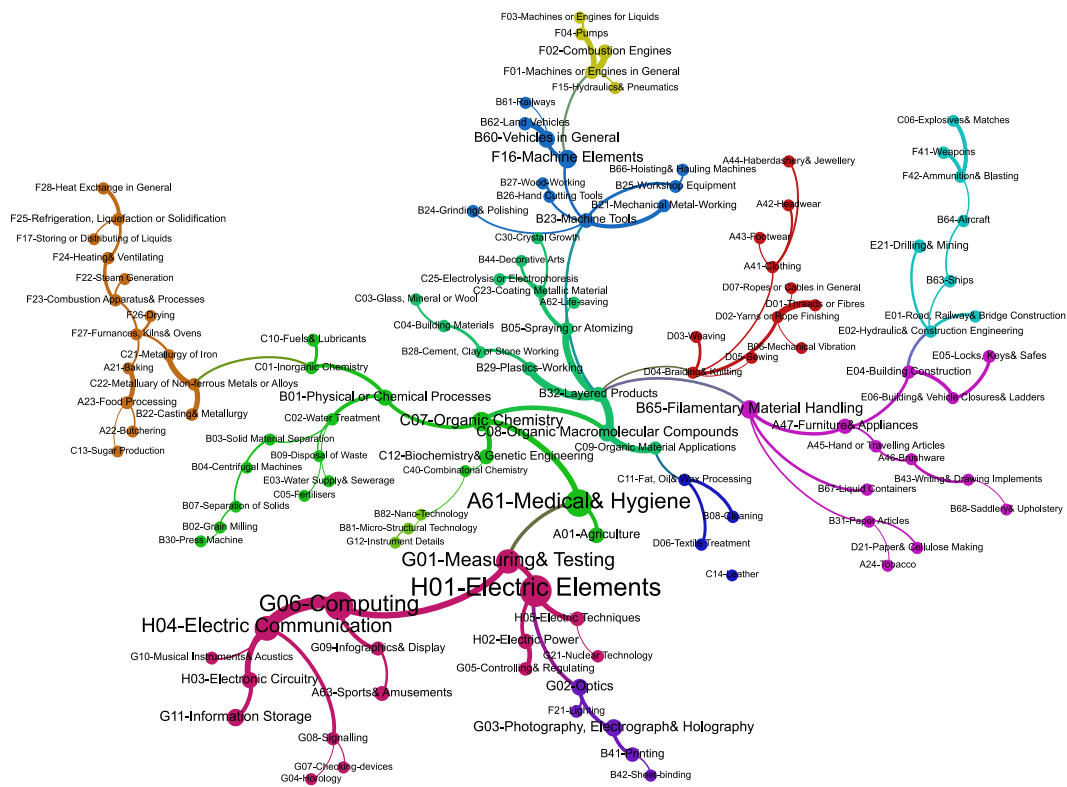

**Figure 5.** Maximum Spanning Tree. Vertex sizes correspond to the numbers of patents in respective IPC patent classes; vertex colors denote communities identified by the Louvain community detection method (Blondel et al., 2008).

Then, we add back to the MST one additional link at a time in the order of decreasing link weight, until reaching the original full network of 7,195 links. Every time a weaker link is added to the network, we calculate the technology network's explanatory power, i.e. Pearson correlation coefficient between the filtered network's link weights and the likelihoods of inventors and organizations diversifying across corresponding links. Figure 6 traces the explanatory power of the technology network for inventor and organization diversification respectively, as we add more and more links back to the MST. The leftmost point of each curve is the MST with 120 links. The rightmost point is the original network that includes all 7,195 links.

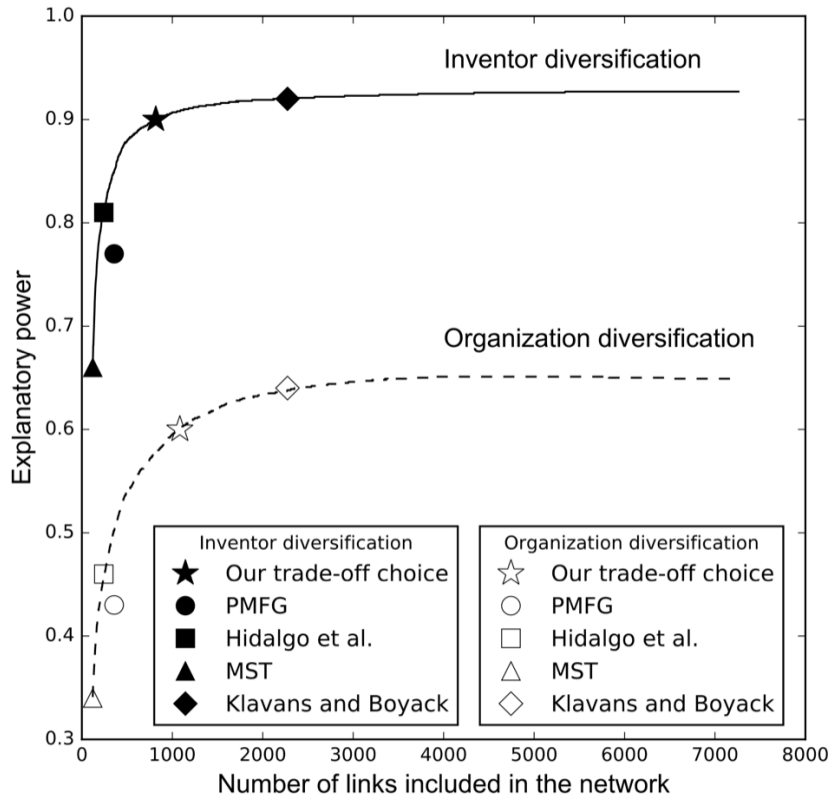

**Figure 6.** The change of explanatory power with the increase of links added to MST.

The shape of these two curves shows the tradeoff between explanatory power and link removal: link reduction incurs the loss of explanatory power, whereas explanatory power increases at the cost of the inclusion of additional links in the network. In particular, both curves present a long flat tail (right segment of the curve) when only the weakest links are removed, and a dramatic dropping (left segment of the curve) when the strongest links are removed. Such a curve shape suggests that diversification explanatory power is insensitive to the removal of thousands of the weakest links, but highly sensitive to the removal of a small number of strong links. This pattern further justifies the filtering strategy of removing weaker links first and strong ones later in order to keep the explanatory power as high as possible.

Figure 6 also reveals that the curve of explanatory power for organization diversification is generally below that for inventor diversification. That is, as discussed earlier, organizations' cross-field diversifications are less constrained, thus less explainable, by knowledge proximity than individual inventors. In addition, the curve shapes are also different for inventors and organizations, i.e., the curve for organization diversification is relatively smoother. Such differences suggest that the thresholds of network filtering need to be decided separately for inventors and organizations.

For the filtered network to explain or predict *inventor diversification* paths, explanatory power increases as more links with lower values are added back to the network (along the upper curve in Figure 6). The left most point, i.e. MST, has the lowest explanatory power of 0.66, indicating that the network filtered at this level does not provide sufficient pathways for inventors to diversify across. The explanatory power climbs up to 0.92 rapidly when the strongest 31.2% of the links (i.e. the strongest 2,242 links) have been included in the network. The highest possible explanatory power is 0.93. After that, the increase in explanatory power with adding more links almost ceased. In other words, adding more links is not useful for increasing the network's explanatory power.

This specific curve shape indicates the filtered network only needs to include a small fraction of strong links to reach a rather high level of explanatory power. For example, to reach a explanatory power of 0.9 (96.8% of the total power), the filtered network only needs to include the 120 MST links plus the additional next-strongest 697 links, totaling 817 links (11.4% of the original network links). This is the "star" point on the upper curve in Figure 6. Figure 7 visualizes this filtered network with 817 links. A vast majority of links has been removed without significantly affecting the network's explanatory power. The structure of the filtered network is more apparent than the full network in Figure 2.

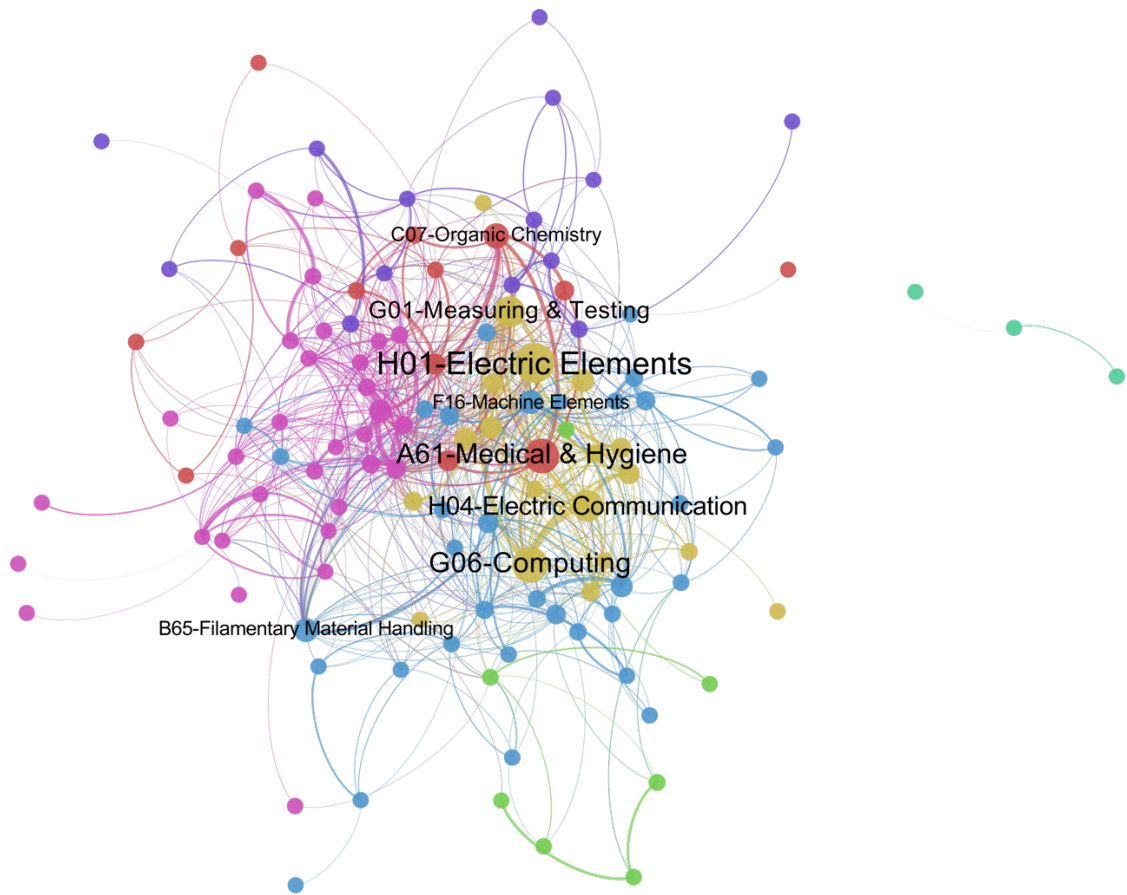

**Figure 7.** Filtered network for explaining inventor diversification paths. The network includes 120 MST links plus additional 697 strongest links of the original network (i.e., 817 links in total), and has a diversification explanatory power of 0.9. Vertex sizes correspond to the numbers of patents in respective IPC patent classes; vertex colors denote communities identified by the Louvain community detection method.

For the filtered network to explain or predict *organization diversification* paths, explanatory power increases initially as more links with lower levels of proximity are added back to MST (along the lower curve in Figure 6). After approximately 61.5% or 4,428 links have been added back to MST, the increase of explanatory power has reached the peak value of 0.651. After the highest point, the explanatory power actually decreases when additional weaker links are included into the network. This suggests that the weakest 38.5% links do not contribute to explanation accuracy, but actually introduce noise to reduce organization diversification explanatory power of the network. Therefore, a filtered network without the weakest 38.5% links can actually have higher explanatory power than the original full network, which was 0.649. This result suggests link removal does not always reduce overall explanatory power of the network, but sometimes improves it.

In general, the specific shape of the curve in Figure 6 indicates that a vast majority of links of the original network can be removed without significantly losing its explanatory power on organization diversification. For example, if one can afford the explanatory power to drop slightly from 0.649 to 0.6, the weakest 6,112 links can be removed from the network. As a result, it only includes the 120 links in MST plus additional 963 strongest links, totaling 1,083 links, i.e., 15.1% of the total original links. This is the "star" point on the lower curve in Figure 6. Figure 8 visualizes this filtered network with 1,083 links for explaining organization diversification, which exhibits a more clear structure than that of the full network in Figure 2. Figure 8 looks similar to Figure 7, because it only adds 266 weaker links.

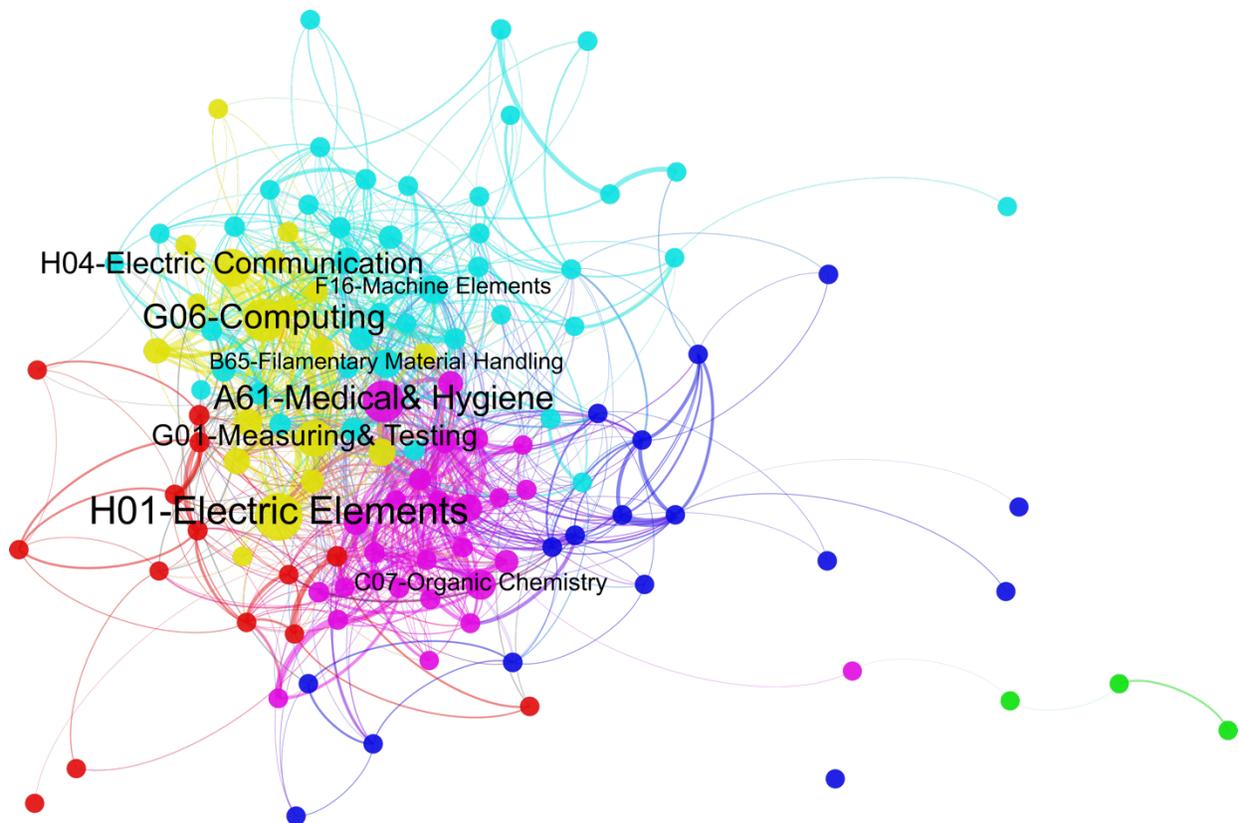

**Figure 8.** Filtered network for explaining organization diversification paths. The network includes 120 MST links plus additional 963 strongest links of the original network, and an explanatory power of 0.60.

Such filtering can be tuned up to any desired level of diversification explanatory power by adjusting the amount of strongest links to add back to the MST. One can make trade-offs between prediction power and link removal, following along the curves in Figure 6. In the following, we compare the diversification explanatory powers of the filtered networks in Figures 7 and 8 with those from alternative network-filtering methods used by Hidalgo et al. (2007) and Klavans and Boyack (2006a) reviewed in section 3, as well as Maximum Spanning Tree (MST) and Planar Maximally Filtered Graph (PMFG) which are two general methods of extracting the most representative set of links from a network. Details of the networks filtered by the above-mentioned methods are as follows:

A) Maximum Spanning Tree (MST) contains the minimal set of the strongest links that keeps all the nodes of the original network connected as a single component. In the context of our paper, the minimum number of links required to connect 121 nodes is 120. The 120 links also have the maximized sum of total weights. Figure 5 is the MST, showing the backbone of the original technology network in Figure 2.

B) Planar Maximally Filtered Graph (PMFG) contains all the strongest links, which can be kept under the constraint of being representable on a plane without any link crossing (planar graph). Tumminello et al. (2005) first introduced PMFG to find the most representative filtered sub-graph of the correlation network of stocks in equity markets, and showed that PMFG always contains the MST as a sub-graph and also preserves the hierarchical organization of MST, while containing a larger amount of links. In the context of our paper, the PMFG of the original technology network includes the strongest 357 links of which the 120 links in MST are a sub-set.

C) Hidalgo et al. (2007) added additional strongest links (in the order of decreasing weights) to the MST, till the point that the number of links is twice of the number of nodes for best visualization, in their search for a good visualization of the product space. In the context of the present paper, we add another 122 strongest links additional to the 120 MST links to result in a network with 242 links and 121 nodes.

D) Klavans and Boyack (2006a) added links back to the network from zero, in the order of decreasing weights, until all unique nodes are connected into the network, in visualizing the networks of academic papers. In the context of the present paper, the network resulted from their approach includes the 2,273 strongest links. Their method is equivalent to removing the weakest links in the order of increasing weights, till the point that the removal of one additional stronger link would cause the network to become disconnected.

Table 2 reports the explanatory powers of the filtered networks resulting from these four alternative filtering methods reviewed above. Only PMFG is below the trade-off curve and clearly not an optimal choice, because many points on the trade-off curve have both better explanatory power and fewer links than the PMFG. For example, the method of Hidalgo et al. (2007) leads to filtered networks that contain fewer links than PMFG, but achieve a higher explanatory power. PMFG's requirement of avoiding link crossing on a plane leads to the inclusion of relatively weaker links, in contrast to Hidalgo et al. (2007) adding links in descending order of weights to MST. MST is the starting point of the trade-off curve, but its explanatory power is the lowest as a result of eliminating too many useful network links.

The network filtered using the method of Klavans and Boyack (2006a) has the highest explanatory powers on both inventor and organization diversification, as it contains many links than MST, PMFG and Hidalgo et al. (2007). One can continue to remove many links at a slight loss of explanatory power. For example, it contains 20% more links to gain a 2% better explanatory power than our network specifically filtered for predicting inventor diversification, and contains 17% more links to gain a 4% better explanatory power than our network specifically filtered for predicting organization diversification. In sum, this network more than doubles the number of links of our two filtered networks to derive only slightly higher explanatory powers. These two filtered networks shown in Figures 7 and 8 lie between Klavans and Boyack (2006a) and Hidalgo et al. (2007) in respective tradeoff curves in Figure 6.

**Table 2.** Comparison of Diversification Explanatory Powers of Different Filtered Networks

| Networks | Number of links after filtering | Power to explain inventor diversification | Power to explain organization diversification |
|---|---|---|---|
| Original network | 7,195 | 0.93 | 0.649 |
| Filtered network for predicting inventor diversification | 817 | 0.90 | N/A |
| Filtered network for predicting organization diversification | 1,083 | N/A | 0.60 |
| A) Maximum Spanning Tree (MST) | 120 | 0.66 | 0.34 |
| B) Planar Maximally Filtered Graph (PMFG) | 357 | 0.77 | 0.43 |
| C) Minimum set of strongest links that contain MST and are 2 x number of nodes (Hidalgo et al., 2007) | 242 | 0.81 | 0.46 |
| D) Minimum set of strongest links to connect all nodes (Klavans and Boyack, 2006a) | 2,273 | 0.92 | 0.64 |

In particular, the curves in Figure 6 suggest no single best threshold but a segment of feasible threshold values to choose from for filtering. Any point in the feasible segment, which has sufficient extent of filtering (i.e., keeping fewer than a threshold amount of links) and high enough explanatory power (i.e., higher than a minimum requirement of explanatory power), can be considered an acceptable threshold for filtering. For example, if one can accept the organization diversification prediction power 0.46, the network of 242 links resulting from the method of Hidalgo et al. (2007) will be acceptable. If one can accept the inclusion of 1,083 links after filtering, the filtered network in Figure 8 will be acceptable and significantly improves explanatory power to 0.6, for organization diversification. In general, with the tradeoff curves, one can more objectively choose a trade-off between explanatory power and the extent of removal of weak links.

## 5. Discussion

We have introduced a method to filter a patent technology network map while maintaining the power of the network map to explain the diversification paths of innovation agents for visualizing such paths of the agents on the map. A few issues related to the method and resultant filtered networks are noteworthy. First of all, different aimed applications of the patent technology map, such as artistic visual effects, may require or favor different filtering approaches and strategies. For instance, these filtering techniques in Table 2 are disadvantaged here for our dual objectives (i.e., fewer links and higher diversification explanatory power), but can be most suitable for the contexts and goals under which they were initially developed.

Second, for our interest in explaining and visualizing diversification paths of innovation agents, strong links are preferred over weak links in the filtering procedure. Prior empirical studies (Breschi et al., 2003; Nesta and Dibiaggio, 2005) have shown that innovation agents are more likely to diversify across technology fields with higher knowledge proximity. That is, stronger links in the network map are more likely to be paths for innovation agents to diversify across. Our Figure 1 shows that, when link weights (measuring knowledge proximity) are higher, link weights are more correlated with diversification likelihoods, indicating a greater contribution of stronger links to the network's overall diversification explanatory power. Figure 6 further shows that the network's diversification explanatory power is insensitive to removing thousands of the weakest links, but drops dramatically when a small number of the strongest links are filtered. These results, together with the empirical literature, all support the preference to keep stronger links and remove weaker links, for our interest in higher diversification explanatory. Similar filtering strategy was commonly adopted in prior patent mapping works (Klavans and Boyack 2006a; Hidalgo et al. 2007; Leydesdorff et al. 2014). In contrast, a different interest in how inventions emerge from combining technologies across distant domains would pay more attention to the weak links.

Furthermore, prior literature on patent technology map construction (Hinze et al., 1997; Yan and Luo, 2016) has shown that the changes of all links' weights and their relative rankings by weights over different years and over even decades are small and insignificant. Weaker links in our network map do not grow stronger and stronger over time. In short, the network of patent technology classes is highly stable overtime. The link weights between different patent classes might be more of the result of the innate knowledge proximity between respective types of technologies represented. Therefore, the stable patent technology network map is unable to directly reveal technology changes. Technology changes might be more effectively investigated via analyzing networks created at a more detailed level, such as patent-to-patent citation network, instead of the network of patent classes that represent technological fields.

For our interest in using the map for capturing and visualizing diversification paths of innovation agents across different technological fields, the stability of the network is beneficial as it justifies the use of overlaying the stable network as a fixed background map with changing technology capability positions (Kay et al., 2014; Yan and Luo, 2016) and diversification paths for a specific innovation agent. In the following, we will demonstrate two different applications examples, one about an individual inventor—Leonard Forbes, and the other about a technology-intensive company—Google.

Note that, different data, parameters and thresholds are used to filter the original technology network, and explain and visualize the diversification paths of inventors vs. organizations. Figure 6 has shown that, given the same extents of filtering of the original network, the filtered network's power of explaining inventor diversification is always higher than that on organization diversification. The calculation of explanatory power for inventor vs. organization diversification requires different information in patent documents, i.e., patenting records of inventors vs. organizations. As a result, the shapes of the "filtering—explanatory power" curves are different for inventors and organizations (Figure 6). Therefore, the thresholds of filtering need to be decided separately for inventors and organizations, resulting in different filtered networks for the applications of inventor and organization diversification analysis.

*5.1 Application Examples*

Figure 9 presents the first example and the diversification paths of Leonard Forbes, who is one of the most prolific inventors of all times and holds more than one thousand patents. We overlay the patent technology network filtered for effectively capturing inventor diversification paths (Figure 7) as the background map, with highlighting the technology fields (i.e., IPC 3-digit classes) where Forbes had patents, and also his most likely paths of patent portfolio diversification over time, based on his patenting history in our database of all USPTO patents till 2010. In contrast, prior patent overlay maps in the literature have not yet visualized the diversification paths of innovation agents (Kay et al., 2014; Hidalgo et al. 2007; Yan and Luo, 2016).

We highlight the *strongest* link to a patent class where Forbes had patents from any patent class where he previously had patents, indicating the most likely path through which Forbes could leverage his prior knowledge to enter a newer technology field.[5] All highlighted arrowed links together constitute 5 maim patent portfolio diversification paths of Forbes, entering 16 technology fields from one to another over time. His main field of patenting is IPC3 class H01 "*electronic elements*", which is also the field where he began his inventing trajectory in 1991. His diversification paths were primarily about information storage and processing, electric and electrical power technologies, optics and material processing.

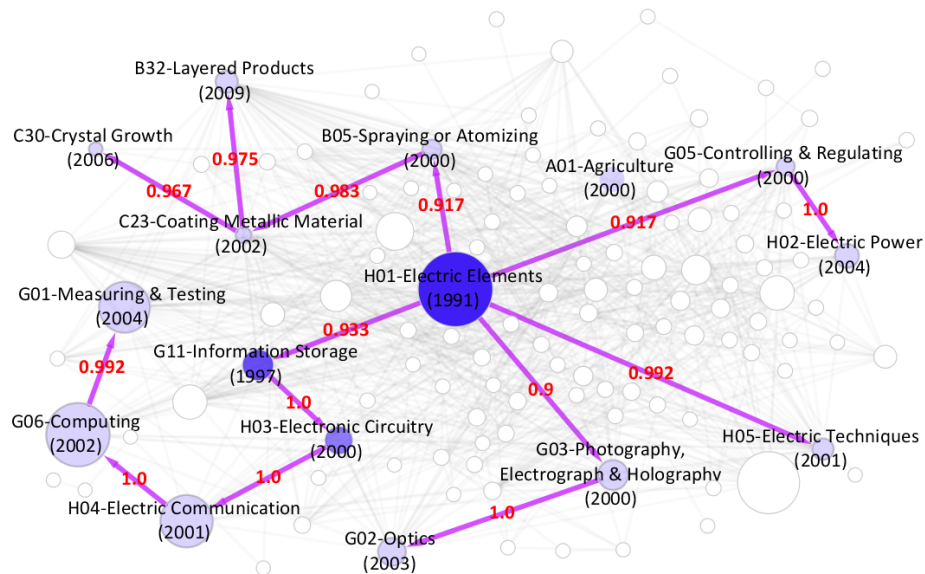

**Figure 9.** Most likely patent portfolio diversification paths of Leonard Forbes. Nodes are IPC classes; node sizes correspond to the number of patents in the represented patent technology classes. Forbes had patents in the highlighted blue nodes; node color intensity denotes the number of patents of Forbes in respective classes. An arrowed link in purple color highlights the strongest link to a field of Forbes from any of the fields that he entered previously. The year when Forbes entered each field is noted in parentheses.

---

[5] To identify the starting field of the strongest link to a newly entered class, in the example we only considered those previously entered fields in which the inventor is still active, e.g. having patents in the most recent 5 years prior to entering the newer target field. If the inventor innovation agent has not had patent in a field for many years since the agent entered field, it indicates that the inventor has left that field and no longer has knowledge or capability there. In both examples in the paper, we also tested 3 years and 7 years for a robustness check. The resulting visualizations are the same.

In the second application example (Figure 10), the agent for diversification analysis is an organization (Google, Inc.). So the background map should be the one filtered for organization diversification (Figure 8). We overlay the background map with the technology fields where Google had patents, and with its most likely paths of diversification over time, based on its patenting records in our database of all USPTO patents till 2010. The overlaid map shows that, Google Inc. (originally founded as a web search engine company) first patented in the field of *computing* (G06), and gradually diversified along 8 main paths into 14 technology fields through 2010.

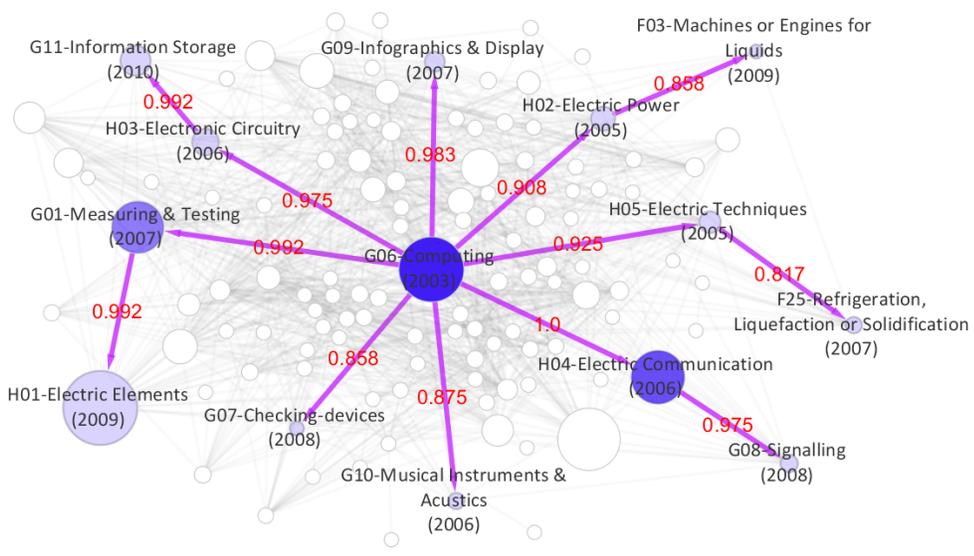

**Figure 11.** Most likely paths of patent portfolio diversification of Google. The nodes are IPC classes; node sizes correspond to the number of patents in the represented patent technology classes. Google had patents in the highlighted blue nodes; node color intensity corresponds to the number of patents of Google in a respective class. An arrowed link in purple color highlights the strongest link to a field of Google from any of the fields that it entered previously. The year when Google entered each field is noted in parentheses.

In one hop from *computing* (G06), Google explored *musical instruments & acoustics* (G10, 2006), *infographics & display* (G09, 2007), and *checking devices* (G07, 2008). The other five paths include two hops in each. One stems from G06 to *electronic communication* (H04, 2006) and then *signaling* (G08, 2008). *Electronic communication* is Google's second most active field after *computing* (G06), reflecting Google's later diversification into smart phone (e.g., Android) and mobile Internet technology. The second two-hop path extends from G06 to *measuring & testing* (G01, 2007) and then *electric elements* (H03, 2009). The third expands from G06 to *electronic* *circuitry* (H03, 2006) and then *information storage* (G11, 2010). The fourth diversified to *refrigeration, liquefaction or solidification* (F25, 2007) via *electric techniques* (H05, 2005). Along the fifth 5-hop path, Google diversified first to *electric power* (H02, 2005) and then to *machines or engines for liquids* (F03, 2009). These paths of patent portfolio diversification reflect Google's initiatives to design data centers of its own, as well as self-driving vehicles.

In both case application examples (Leonard Forbes and Google), these links in the most likely diversification paths are also the strongest ones among all links of their starting fields, as indicated by their high *p* values (reported on respective links in the maps), i.e., the percentile of the starting field's all links which have lower or equal link values than the arrowed link from the starting field.[6] The overall high *p* values for all outgoing links of starting fields indicate that the most likely unexplored fields to enter next are those with the highest proximities to any of the previously entered fields. Following this, one can further extend from past positions and trajectories of an innovation agent to predict its future fields to enter next on the map. In particular, because the background network is rather stable even over decades, the same map can be used to identify most feasible future fields for an innovation agent to enter in a reasonably short term future. In brief, if the network could well explain the historical paths of diversification of an innovation agent (such as the two examples above), it should also be able to predict the "most likely" future fields at least in a near future.

## 6. Summary

In this paper, we have introduced a method to filter a patent technology network map for visualizing diversification paths of innovation agents. Patent technology network contains many weak links, which do not contribute to accuracy and might add noise to diversification path analysis and complexity in visualization. But removing links from the network may reduce the power of the filtered network to predict innovation agents' technology diversification across fields. Despite this conflict, because link weight distribution is highly skewed, most links are extremely weak and can be removed without significantly affecting the network's diversification explanatory power.

---

[6] For example, the *p* value of 0.9333 for the link from *electric elements* (H01, 1991) to *information storage* (G11, 1997) means that the value of this link is greater than or equal to 93.33% of all the non-zero links of class H01.

Specifically, our method is to explore the curve of the explanatory power of filtered network against the number of strongest links included in the network, and stop at a threshold, which contains a sufficiently small number of strong links to achieve a sufficiently high explanatory power. Our main contribution is the trade-off curve between link removal and diversification prediction power, which provides an objective basis to rule out the obviously non-optimal choices, such as PMFG, and to compare and select feasible thresholds in the center segment of the curve. To obtain the trade-off curve, we also proposed a new metric of such explanatory power. Our filtering method is fundamentally different from those from the literature in that it aims to maintain the effectiveness of using the filtered network to explain and visualize the technological or patent portfolio diversification paths of innovation agents.

Moving forward, the technology networks can be filtered for the interests of additional types of innovation agents. For instance, one can use it to analyze the diversification paths of regions and countries, in addition to individuals and organizations. One can also use other categories or aggregations of patents (Benson and Magee, 2013; 2015; Kay et al., 2014) than the 3-digit IPC classes to create technology network maps, and investigate if they are more meaningful. In the present paper, we only used one single metric to quantify explanatory power. It will be meaningful to explore additional metrics, and test if the results hold.

Furthermore, the analyses in the present paper, including the two application examples, were about describing the history. The use of the map can be extended for foresight analyses—identifying the future fields of an innovation agent, based on assessing the unexplored fields in terms of their proximities to its previously entered fields. In general, there is great potential to apply the filtered technology network map to aiding in capability building trajectory analyses, technology road mapping, and the exploration of innovation directions of different kinds of innovation agents.

**Acknowledgement**

This research is supported partly by a grant from the SUTD-MIT International Design Centre (IDG31300112), and by Singapore Ministry of Education Tier 2 Academic Research Grant (T2MOE1403).